\documentclass[conference]{IEEEtran}
\IEEEoverridecommandlockouts
\usepackage[colorlinks,linktoc=all]{hyperref}
\usepackage[all]{hypcap}
\hypersetup{citecolor=MidnightBlue}
\hypersetup{linkcolor=MidnightBlue}
\hypersetup{urlcolor=MidnightBlue}
\usepackage[nameinlink]{cleveref}
\creflabelformat{equation}{#2\textup{#1}#3}  

\usepackage{amsmath}
\usepackage{amsthm}
\usepackage{amssymb}

\usepackage{color}
\usepackage[usenames,dvipsnames]{xcolor}
\definecolor{shadecolor}{gray}{0.9}

\usepackage{grffile}
\usepackage{graphicx}

\graphicspath{{images/}}

\usepackage{booktabs, array, tabulary}

\newcommand{\colr}[1]{{\color{red} #1}}
\newcommand{\colb}[1]{{\color{blue} #1}}

\begin{document}

\title{An industry case of large-scale demand forecasting of hierarchical components\\
\thanks{Sections I, II, III, IV were supported by the Ministry of Education  and  Science  of  the  Russian  Federation  (Grant no. 14.756.31.0001). Other sections were supported by the Mexican National Council for Science and Technology (CONACYT), 2018-000009-01EXTF-00154. The authors would like to thank Huawei Noah's Ark Lab for the advice and support.}
}
\author{
    \IEEEauthorblockN{
    Rodrigo Rivera-Castro\IEEEauthorrefmark{1}, 
    Ivan Nazarov\IEEEauthorrefmark{1}, 
    Yuke Xiang\IEEEauthorrefmark{2}, 
    Ivan Maksimov\IEEEauthorrefmark{1},
    Aleksandr Pletnev\IEEEauthorrefmark{1}
    Evgeny Burnaev\IEEEauthorrefmark{1}}
    \IEEEauthorblockA{\IEEEauthorrefmark{1}Skolkovo Institute of Science and Technology
    \\rodrigo.riveracastro@skoltech.ru}
    \IEEEauthorblockA{\IEEEauthorrefmark{2}Huawei Noah's Ark Lab}
}

\maketitle

\begin{abstract}
Demand forecasting of hierarchical components is essential in manufacturing. However, its discussion in the machine-learning literature has been limited, and judgemental forecasts remain pervasive in the industry. Demand planners require easy-to-understand tools capable of delivering state-of-the-art results. This work presents an industry case of demand forecasting at one of the largest manufacturers of electronics in the world. It seeks to support practitioners with five contributions: (1) A benchmark of fourteen demand forecast methods applied to a relevant data set, (2) A data transformation technique yielding comparable results with state of the art, (3) An alternative to ARIMA based on matrix factorization, (4) A model selection technique based on topological data analysis for time series and (5) A novel data set. Organizations seeking to up-skill existing personnel and increase forecast accuracy will find value in this work.
\end{abstract}


\section{Originality and Value}
This research presents a demand forecasting system of electronic components in manufacturing validated with real data. The contributions cover the areas of pre-processing, prediction and model selection and are suited for individuals with domain knowledge but limited understanding of machine learning methods. They are the following:
\begin{enumerate}
\item An industry case of demand prediction for a large manufacturer of electronics,
\item An evaluation of 14 different models for demand prediction of items with hierarchical dependencies,
\item An implementation of a method for demand forecasting based on matrix factorization,
\item A feature engineering technique that is both easy to implement and yields similar results to those obtained from using feature engineering requiring domain knowledge,
\item A methodology for model selection based on topological data analysis suited for large data sets in an industry-setting,
\item For reproducibility purposes, an implementation, and data set available for download\footnote{\url{https://github.com/rodrigorivera/icmla2019}}.
\end{enumerate}

\section{Problem Statement}\label{sec:problemStatement}
One of the world's largest manufacturers of electronics has to forecast demand for both its products and their respective individual components, amounting to millions of time series data to predict. Traditional forecasting techniques are here largely ineffective. Nevertheless, the manufacturer has to generate reliable estimates for its future demand over multiple periods.

\section{Research Abstract and Goals}\label{researchandgoals}
At the moment, there are more than \$12 trillion USD in inventory either stockpiled or in transit, amounting to 17\% of the world's Gross Domestic Product (GDP), \cite{doi:10.1080/00207543.2018.1524167}. An accurate demand forecasting is essential in the industry. 
Nevertheless, imprecise demand planning is still pervasive. For new products, forecast errors are, on average 44-53\%, whereas, for improved products, it is 31\%, \cite{Kahn2000AnEI, jain2005}. Companies compensate for this inaccuracy gap through expensive operational measures such as trans-shipments \cite{simchi2008designing}. Nevertheless, retailers still experience out-of-stock (OOS) events with rates amounting to 8.3\% worldwide, \cite{gruen2002retail}.

\begin{figure}[!ht]
  \begin{center}
    \includegraphics[width=\columnwidth]{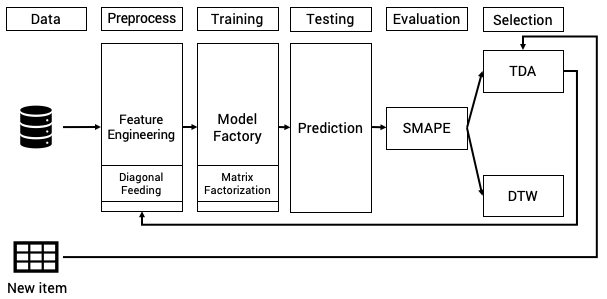}
  \end{center}
  \caption{Overview of the implemented methodology}
  \label{fig:intro:system_overview}
\end{figure}

The objective of this research is to present three techniques for (1) data pre-processing, (2) prediction, and (3) model selection accessible to non-technical business experts and offering competitive results. They represent a cohesive system depicted in \autoref{fig:intro:system_overview}. The use of novel machine learning methods for this field is a promising area with little academic research and with insufficient efforts to expose practitioners to them, \cite{2017arXiv170905548R}. It is relevant to have robust methods accessible to broader audiences, \cite{Chase2013}. \cite{Fleisch2003} observed that for discrepancies as low as 2\%, it is worth investing in improving the accuracy of a forecast. \cite{wipro2013} goes as far as claiming that a 10\% reduction in OOS increases revenue of retailers by up to 0.5\%.
Nevertheless, companies struggle to hire adequate personnel to address these tasks, \cite{2017apec}. \cite{2018esade} reported that over 60\% of surveyed businesses are resorting to internal training to compensate for this. This work seeks to alleviate this situation by presenting an extensive comparison of methods, proposing a feature engineering technique well-suited for demand forecasting in manufacturing, evaluating a novel method based on matrix factorization, and proposing a technique for model selection that is both accurate as well as easy to communicate to decision-makers.
The research goal of this work is to propose a set of approaches for time series forecast that can be adopted by business practitioners. For this purpose, the study poses the questions: 1) What is state of the art in academic research of time series prediction with structures? 2) How does the proposed method differs from popular approaches applied to time series prediction tasks? 
Two objectives achieve the research goal: a) To review the existing theory on time series prediction and especially on techniques for dynamic hierarchical structures; b) To make a performance comparison of the proposed technique.
The object of research is the balance between accessibility and precision of methods for time series in a massive data context within the industry. 
The subject of the research is forecasting product demand using techniques for time series with hierarchies.

\section{Literature Review}\label{section:literaturereview}
Supply chain management (SCM) in general and demand forecasting, in particular, are fields that have commanded attention from different communities according to \cite{Attar2016}. A comprehensive treatment is available in the works of \cite{Chase2013} and \cite{Gilliand2015}. Sales forecasting is an essential part of the supply chain management. The forecasting community uses and trains quantitative methods of the statistical family of ARIMA, exponential smoothing models, and alike with historical data to forecast future points to improve the forecasting accuracy. However, \cite{Ahmed2010} argues that there have been few large scale comparative studies of machine learning models for regression or time series aimed at forecasting problems. 
In the retail and manufacturing sectors, authors such as \cite{Tirkes2017} paid attention to the demand forecasting of edibles.
Similarly, \cite{Taylor2007} deals with time series characterized by a high volatility skewness to forecast daily sales for a supermarket chain at the point of sale. 
\cite{Bianchi2017} experimented with recurrent neural networks for short term forecasting of real-valued time series. While \cite{Carbonneau2008} explored demand forecasting with incomplete information. 
For the electronics manufacturing industry, \cite{Wan2016} introduced SVM regression to the supply chain of various producers. 
Although SVM regression is a popular method for forecasting, not everyone has identified it as the most effective method. For example, \cite{Lu2012} presented a MARS model, and \cite{Yelland2010} proposed a Bayesian model.
Other manufacturing-centric sectors such as fashion have also delved into demand forecasting but to a different extent. \cite{Liu2013} claims that pure statistical methods are not yet commonplace in the fashion industry. It is preferred to make use of judgmental forecasts or a combination of quantitative and qualitative forecasts.

\section{Dataset}
\begin{figure}[!ht]
  \begin{center}
    \includegraphics[width=0.9\columnwidth]{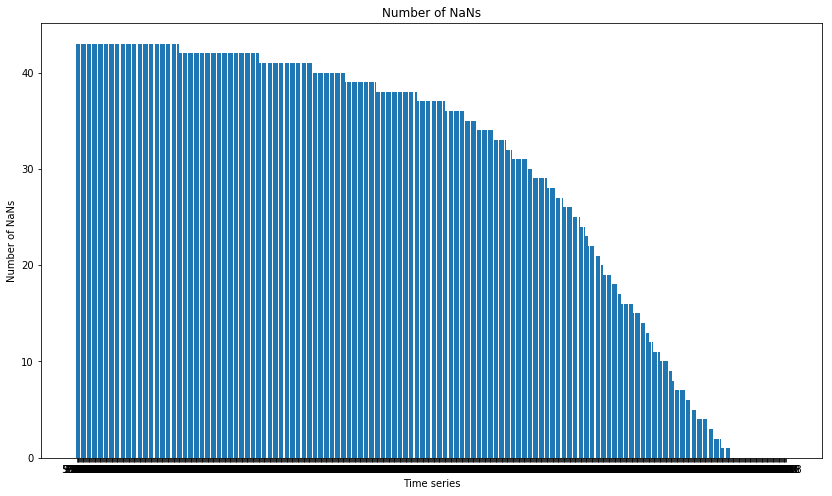}
  \end{center}
  \caption{Number of NaNs (zero orders) per item. X-axis: Item's ID, Y-axis: Number of zeroes}
  \label{fig:tda:orders per item:2}
\end{figure}

The data consists of a data set of observations from an electronics manufacturer representing a subset of their total inventory. It contains the demand for 2562 different items with a length of $n = 45$. These items have varying amounts of required quantities, with many of them being requested sporadically, as seen in \autoref{fig:tda:orders per item:2}, and few of them being requested in large quantities.

\section{Diagonal feeding} 
\label{sec:diagonal_feeding}
One of the contributions of this work is to introduce the practitioner to a data transformation technique, useful for multi-step structured forecasting from anticipatory data. It is part of the first step, 'Preprocess,' of the system introduced in \autoref{fig:intro:system_overview}. The main benefit of Diagonal Feeding is that it helps utilize the anticipatory nature of pre-orders' time-series data and makes forecasting the pre-order structure more streamlined.
It is made possible due to the data set containing information not only about the current demand but also on the volumes of pre-orders made in advance. Advance pre-orders are expectation-driven, naturally forward-looking, and known beforehand, as they reflect planning and anticipation of the market and economic environment at the end of the period when the order is to be fulfilled. At the same time, forecasting the pre-order structure over several next periods is of significant practical interest. It is reasonable to leverage the anticipatory information of the advance pre-orders, known by the present, for predicting the pre-order structure in the future by also taking into account the cross-correlations between the pre-orders.
Let $q_t^h$ be the volume of some item in the ``quantity'' field in the data set, $t$ corresponds to the ``delivery date'', and $h$ be the ``periods before delivery date''. The value $q_t^h$ denotes the total amount requested via {\it $h$ period advance pre-orders} to be delivered {\it by the end of period $t$}. The key property of the data set is that for every item, the value $q_t^h$ is effectively-known and available for use by the end of period $t - h$ -- the period when the $h$-ahead pre-orders were made. For example, $q_{t-1}^1$ is known at the end of $t-2$. It corresponds to the quantity requested at the end of $t-1$. That is the case due to the accumulation of pre-orders made by the end of $t-2$.
Since $q_t^h$ reflects expectations about the market conditions at $t$ and is known $h$ periods in advance, it seems reasonable to reorder the data set with respect to the period when they become known and reshape it to keep the pre-order structure. This makes predicting $q_t^h$ with $q_s^f$ data for $t - h > s - f$, which is either past ($q_{t-1}^h$) or anticipatory ($q_t^{h+1}$), more streamlined. 
The proposed reshaping of the multivariate time series of a particular ``item'' is illustrated below. Since the quantity $q_t^h$ is known at time $t-h$, each diagonal $(q_{t+s+h}^h)_{h\geq 0}$ in the scheme above is {\it known} at $t+s$, $s\in \mathbb{Z}$; thus, potentially up to infinite periods:
\begin{small}
\begin{equation} \label{eq:diag_feed}
\begin{aligned}
  \begin{pmatrix}
      \colb{q_{t+0}^0} & \colb{q_{t+0}^1} & \colb{q_{t+0}^2} \\
      \colr{q_{t+1}^0} & \colb{q_{t+1}^1} & \colb{q_{t+1}^2} \\
      \colr{q_{t+2}^0} & \colr{q_{t+2}^1} & \colb{q_{t+2}^2} \\
      \colr{q_{t+3}^0} & \colr{q_{t+3}^1} & \colr{q_{t+3}^2} \\
  \end{pmatrix}
  \rightarrow
  \begin{pmatrix}
      \colb{x_{t 0}} & \colb{x_{t 1}} & \colb{x_{t 2}} \\
      \colr{y_{t 0}} & \colb{x_{t 3}} & \colb{x_{t 4}} \\
      \colr{y_{t 1}} & \colr{y_{t 2}} & \colb{x_{t 5}} \\
      \colr{y_{t 3}} & \colr{y_{t 4}} & \colr{y_{t 5}} \\
  \end{pmatrix}.
  \end{aligned}
\end{equation}
\end{small}
In \autoref{eq:diag_feed}, the target $\colr{y_t}$ is the output and represents the pre-order structure for the next $3$ periods beginning with $t+1$. The objective is to predict the lower diagonal of the matrix. It is done using $\colb{x_t}$ and its history as an input, i.e., past pre-order structure. Although, in principle, predicting the structure in $\colr{y_t}$ allows planning production volumes several months ahead, the most relevant targets for practical demand forecasting are on the largest diagonal of $\colr{y_t}$, since they are the earliest future volumes.

\section{Matrix Factorization}
\label{sec:matrix_factorization}
Matrix Factorization (MF) methods are used in a variety of applications such as recommender systems, signal processing, \cite{Weng2012}, computer vision, \cite{Chen2004}, and others. The second contribution of this work is adapting a method discussed in \cite{Rivera_2018} and \cite{yuetal2016} to demand forecasting in manufacturing.
Let $Y$ be $T\times n$ sparse or dense matrix of observations of $n$ objects spanning the period of $T$ time steps, i.e. each column $i=1,\,\ldots,\,n$ of $Y$ is a times series $y^{(i)} = (Y_{ti})_{t=1}^T$ related to the $i$-th object. The problem of factorizing a fully or partially observed $T\times n$ matrix $Y$ consists of finding $d$-dimensional factors $Z$ and the corresponding factor loadings $F$. It must be in the form of $T \times d$ and $d \times n$ matrices respectively. As such, their product $Z F$ most accurately recovers the observed $Y$, i.e. $Y_{ti} \approx \sum_{j=1}^d Z_{tj} F_{ji}$. This is usually achieved by solving the following optimization problem:
\begin{equation} \label{eq:general_mf}
\begin{aligned}
& \underset{F, Z}{\text{minimize}}
  & & \tfrac1{2 \lvert \Omega\rvert}
        \|\mathcal{P}_\Omega(Y - Z F)\|^2
      + \lambda_F \mathcal{R}_F(F)
      + \lambda_Z \mathcal{R}_Z(Z)
      \,,
\end{aligned}
\end{equation}

where $\Omega\subset \{1..T\} \times \{1..n\}$ is the sparsity pattern of $Y$, $\mathcal{P}_\Omega$ zeroes out unobserved entries. The coefficients $\lambda_F$ and $\lambda_Z$ are non-negative regularization coefficients that govern the trade-off between the reconstruction error and the regularizing terms $\mathcal{R}_F$ and $\mathcal{R}_Z$. The latter depends on the particular desired properties of the factorization, typically in conjunction with a Ridge regression-type penalty ($\ell^2$ norm).

\section{Models}\label{sec:models}
The third contribution of this work is a large-scale study of various methods for demand forecasting. In the system presented in \autoref{fig:intro:system_overview}, they belong to the parts 'Training' and 'Testing.' In total, the assessment consists of fourteen different methods. They are 1) Adaboost, 2) ARIMAX, 3) ARIMA, 4) Bayesian Structural Time Series (BSTS), 5) Bayesian Structural Time Series with a Bayesian Classifier (BSTS Classifier), 6) Ensemble of Gradient Boosting (Ensemble), 7) Ridge regression (Ridge), 8) Kernel regression (Kernel), 9) Lasso, 10) Matrix Factorization from \autoref{sec:matrix_factorization} (MF), 11) Neural Network (NN), 12) Poisson regression (Poisson), 13) Random Forest (RF), 14) Support Vector Regression (SVR).
Each of them had as a target value three different options: a) Quantity (non-transformed), b) Log-transformed quantity, c) Min-Max transformed quantity. Additionally, Diagonal Feeding, presented in \autoref{sec:diagonal_feeding}, was evaluated for regression methods. Thus, one evaluates three settings: a) No Diagonal Feeding, b) Diagonal Feeding with an item by item training (One by One). In this case, a vector containing the input of a specific item is fed individually to a model, c) Diagonal Feeding fitting the model on the full data set (All Items). Here, one uses a matrix with the input from all items. In all three cases, one obtains an individual vector corresponding to a given item as an output. For a), extensive feature engineering is necessary, and the outcome was 360 features.
The specific features are documented in the code base provided\footnote{\url{https://github.com/rodrigorivera/icmla2019}}. The training set consisted of 37 periods, and the test set of 8. The Symmetric Mean Absolute Percent Error (SMAPE) serves to evaluate the performance of the models, and one defines it as $
\text{SMAPE} = \frac{200\%}{n} \sum_{t=1}^n \frac{|F_t-A_t|}{|A_t|+|F_t|}
$ 
with $F_t$ being the forecasted value and $A_t$ the actual value at time $t$ respectively. One can see the results of the experiment in \autoref{tab:models:2}. The table contains both the median and average SMAPE for all models, an average for models fit without Diagonal Feeding (DF), and a second average in the case where it was used.  Further, the performance across models was uneven. The top 5 of models that achieved the lowest SMAPE for a given item were 1) Adaboost with 222 items, 2) Ensemble of Random Forests with 45, 3) BSTS with 42, 4) BSTS Classifier with 32 and 5) ARIMAX with 21 respectively.

\begin{small}
  \begin{table}
\caption{Overview of results using mean SMAPE. Low values are better. 1:1: One by One. AI: All Items. MM: Min-Max. LT: Log-Transform. DF: Diagonal Feeding}
\begin{center}\label{tab:models:2}
\begin{tabulary}{\linewidth}{CCCC}
    \toprule
Model	&	SMAPE	&	Model	&	SMAPE	\\	\hline
Adaboost	&	0,17	&	Ridge 1:1 MM DF	&	0,42	\\
Ensemble	&	0,18	&	Adaboost 1:1 LT DF	&	0,43	\\
ARIMA	&	0,27	&	Kernel AI LT DF	&	0,43	\\
Ridge	&	0,3	&	Ridge AI MM DF	&	0,43	\\
SVR	&	0,3	&	Kernel 1:1 DF	&	0,44	\\
ARIMAX	&	0,32	&	Adaboost 1:1 DF	&	0,47	\\
RF 1:1 LT DF	&	0,34	&	Adaboost 1:1 MM DF	&	0,47	\\
Poisson AI LT DF	&	0,36	&	Kernel 1:1 LT DF	&	0,47	\\
Lasso AI DF	&	0,37	&	NN AI MM DF	&	0,47	\\
Poisson 1:1 DF	&	0,37	&	MF	&	0,5	\\
Poisson 1:1 LT DF	&	0,37	&	NN 1:1 MM DF	&	0,52	\\
Poisson 1:1 MM DF	&	0,37	&	\textbf{AVERAGE ALL}	&	0,53	\\
RF 1:1 DF	&	0,37	&	\textbf{AVERAGE DF}	&	0,54	\\
RF 1:1 MM DF	&	0,37	&	SVR 1:1 LT DF	&	0,55	\\
Ridge AI DF	&	0,37	&	Adaboost AI LT DF	&	0,56	\\
Lasso 1:1 DF	&	0,38	&	SVR AI LT DF	&	0,56	\\
NN AI LT DF	&	0,38	&	SVR 1:1 DF	&	0,56	\\
RF AI LT DF	&	0,38	&	Lasso 1:1 MM DF	&	0,6	\\
Ridge 1:1 DF	&	0,38	&	SVR 1:1 MM DF	&	0,6	\\
Kernel AI DF	&	0,39	&	RF AI DF	&	0,62	\\
Kernel AI MM DF	&	0,39	&	RF AI MM DF	&	0,62	\\
Lasso 1:1 LT DF	&	0,39	&	NN 1:1 DF	&	0,68	\\
Poisson AI DF	&	0,4	&	NN 1:1 LT DF	&	0,74	\\
Poisson AI MM DF	&	0,4	&	SVR AI DF	&	0,87	\\
Ridge 1:1 LT DF	&	0,4	&	BSTS	&	0,93	\\
Lasso AI LT DF	&	0,41	&	BSTS classifier	&	0,97	\\
Ridge	&	0,41	&	Adaboost AI DF	&	1,1	\\
Ridge AI LT DF	&	0,41	&	Adaboost AI MM DF	&	1,11	\\
\textbf{MEDIAN ALL}	&	0,42	&	NN AI DF	&	1,13	\\
\textbf{AVERAGE NO DF}	&	0,42	&	Lasso AI MM DF	&	1,15	\\
Kernel 1:1 MM DF	&	0,42	&	SVR AI MM DF	&	1,76	\\
 \bottomrule
 \end{tabulary}
 \end{center}
 \end{table}
 \end{small}

\section{TDA for Model Selection}\label{sec:tda}
\begin{figure}[!b]
  \begin{center}
    \includegraphics[width=0.9\columnwidth]{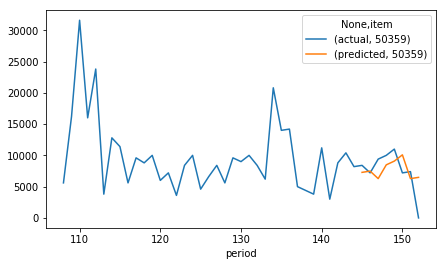}
  \end{center}
  \caption{Top forecast using ARIMA. X-axis: Period. Y-axis: Quantity. Blue color: Actual quantity. Orange color: Predicted quantity. SMAPE: 0,20}
  \label{fig:models:autoarima:1}
\end{figure}

In the system presented in \autoref{fig:intro:system_overview}, model selection is done with a method based on Topological Data Analysis. It represents the fourth contribution of this study. TDA is a new field that emerged from a combination of various statistical, computational, and topological methods during the first decade of the century. It allows us to find shape-like structures in the data and has proven to be a powerful exploratory approach for noisy and multi-dimensional data sets. For a detailed introduction, the reader is invited to consult \cite{chazal2017introduction}. 
Two motivations lie behind this approach. First, in a production-setting with millions of time series to forecast, it is necessary in advance to decide on the appropriate model for a particular item in order to minimize computing costs and efforts. There are many periods with zero orders and peaks in demand. Second, SMAPE as the sole metric for decision-making can be misleading, especially if it is evaluated exclusively on the training set. For example in \autoref{fig:models:autoarima:1}, the best forecast using ARIMA is depicted. A relatively low SMAPE of 0,20 was obtained. Nevertheless, the model is only predicting the value at time $t+1$ using the value from time $t$.

\begin{figure}[!b]
  \begin{center}
    \includegraphics[width=0.7\columnwidth]{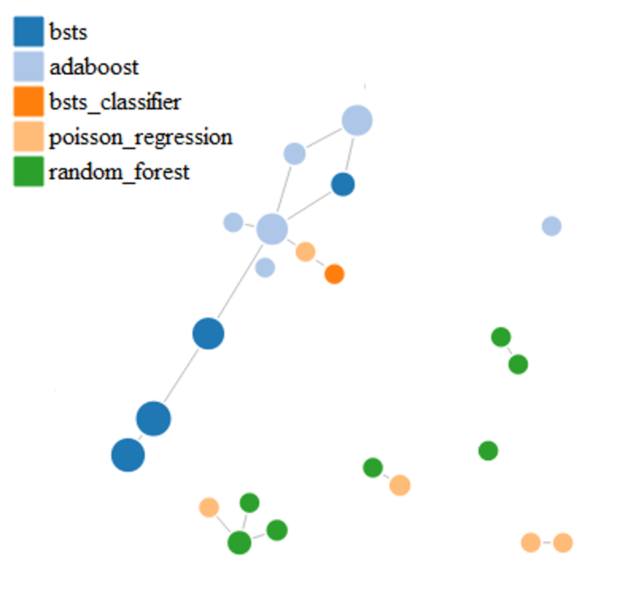}
  \end{center}
  \caption{TDA pipeline for 5 models and 7 features with Canberra distance. Colors: Blue (BSTS), Orange (BTSTS classifier), Yellow (Poisson), Green (RF), Grey (Adaboost).}
  \label{fig:tda:results}
\end{figure}

This research proposes a pipeline consisting of 8 steps to select a model. (A) For a subset of time series, in this case, 200, all possible models are fitted. For this experiment, one used only five models, see \autoref{fig:tda:results}. (B) On the test dataset and for the same items, one calculates SMAPE for each model. (C) For each time series, the best model is chosen based on SMAPE. The best model becomes a target. (D) One computes relevant features describing each time series, see \cite{christ2018time}. (F) A graph is constructed using the Mapper algorithm, see \cite{chazal2017introduction}. The Canberra distance, see \cite{lance1967mixed}, is used as a distance metric and the first principal component obtained from the Mapper algorithm as a lens.
(G) A graph partitioning algorithm, see \cite{slininger2013fiedlers}, is run recursively until reaching the lowest limit of data points per cluster. (H) One chooses the most frequently observed target (model) for all models within a cluster. (I) For a new time series, one can select the best model by running the K-nearest neighbors algorithm on the features obtained in point (D). For this experiment, one chooses seven features.
Based on the described pipeline, one obtains two clusters of nodes from the graph: a) AdaBoost, BSTS, BSTS classifier for 74\% of the time series, b) Poisson regression, and Random Forest for 26\% respectively. They are depicted in \autoref{fig:tda:results}. Using cross-validation for model selection, for 71\% of the time series, AdaBoost, BSTS, BSTS Classifier were the best choice. Hence, using only one graph, partitioning yields a small model selection error (6\%).

\section{Discussion \& Learnings}
\paragraph{On Diagonal Feeding}\label{sec:discussion:df}
The critical insight from the analysis of the data set through Diagonal Feeding is that the currently known one-period mostly determines the next period's gross total demand volume $q_{t+1}^0$ ahead pre-orders for the period ($q_{t+1}^1$). 
The net-next period's volume, $\delta_{t+1}^0$, is the difference between $q_{t+1}^0$ and $q_{t+1}^1$. Viewed through Diagonal Feeding, it is mostly independent of the history of net pre-orders for the period $t+1$ and is thus less predictable from advance pre-order data, as indicated by the correlation analysis and the results of a grid search experiment. The apparent success of forecasting the $q_{t+1}^0$, especially in contrast to the other next period's pre-order volumes $q_{t+1+j}^j$ for $j \in \{1,2,3\}$, might be attributed to an observed high correlation of the one-period ahead pre-order volume $q_{t+1}^1$. Further, Diagonal Feeding delivers results comparable to those obtained doing extensive feature engineering. Along these lines, exploring different transformations of the target value is essential. For example, a Neural Network without a transformed quantity fitted on the full data set had a SMAPE of 1,13, with a log transformation, it was 0,38.

\paragraph{On Matrix Factorization}\label{sec:learnings:mf}
The contribution of this work concerning \cite{yuetal2016} is an implementation of MF with temporal regularization solving explicitly the following optimization problem (extended with graph similarity regularizer).
The major advantage is that in the high dimensional object mode $T \ll n$, it has fewer parameters ($T k + k n  + k p$) to estimate than $p$-th order vector autoregression ($p n^2$), while retaining the power to capture the correlations among the time series in $Y$,~\cite{yuetal2016}. Nevertheless, criticism is twofold. First, the method is wasteful. Its most precise forecasts are one-step-ahead, since it relies on the ``dynamic'' forecasting method: the factor forecasts are computed based on the prior forecasts $\hat{Z}_{T+h\mid T,\, j}    = \sum_{i=1}^p \phi_{ji} \hat{Z}_{T+h-i\mid T,\, j}$ with $\hat{Z}_{T+h-i\mid T,\, j} = Z_{T+h-i,\, j}$ for $i\geq h$. One attributes this deterioration of forecast accuracy to the accumulating forecast error inherent to this method. The secondary reason is that the $\ell^2$ and $\mathrm{AR}(p)$ regularizers jointly force stationary factor time series $(Z_{t,\,j})_{t=1}^T$, with the characteristic roots lying within the $\mathbb{C}$ unit disk. Therefore, the dynamic forecast, although capable of exhibiting complex dynamic patterns for high $p$, still has vanishing oscillations, eventually leveling to zero. The second shortcoming is that it is impossible to get the new latent factor values when one updates $Y$ with new data, other than re-estimating the factorization model. The key issue with re-estimation is that the re-estimated factors and loadings are not guaranteed to resemble the ones from the factorization before the data update.
Given these shortcomings, following guidelines for the application of the temporal regularized matrix factorization can be formulated. First, one should observe the experiments by \cite{yuetal2016} suggest that at least $25\%$ of the entries in $Y$ for an adequate reconstruction of the missing dynamics within the training set. Second, the structure of the $\mathrm{AR}(p)$ regularizer suggests that the factorization should not express extreme volatility. A comparison of the performance of this method with the second data set (non-sparse and moderately volatile) against the third one (highly sparse and volatile) supports this. Third, due to the dynamic nature of the factor forecasts, the best strategy is to compute one-step-ahead forecasts and re-estimate the factorization upon new data.

\paragraph{On the experiment}
The results from \autoref{tab:models:2} show that the best model was Adaboost with an SMAPE of 0,17. It was followed by the Ensemble of Random Forests with 0,18. Both performed significantly better than Arimax, the baseline used by the manufacturer, with 0,32. Worth highlighting are the results obtained by Diagonal Feeding. The best method using this transformation technique, a random forest with log-transform and fitted on the full data set, obtained 0,34. It was significantly better than an average consisting of methods trained on 360 features with a SMAPE of 0,42.

\paragraph{On the scope of the study}
The objective of this study was to improve the results obtained from the forecast method used by the manufacturer, ARIMAX.  At the same time, it seeks to provide tools that demand planners at the electronics manufacturer can use without requiring extensive knowledge in computer science. In this study, ARIMAX showed good results using SMAPE as an error metric. However, looking at individual items, it only gave the best results for less than 10\% of the inventory. Besides, it showed that using Diagonal Feeding improves results without extensive feature engineering. From an academic perspective, this study filled a void. In the literature, there are no comprehensive studies on demand forecasting for manufacturers that practitioners can use as a reference.

\paragraph{On TDA for Model Selection}
The manufacturer's inventory consists of millions of components. Thus, proper and efficient model selection becomes essential. Model selection based on TDA produced fast and explainable results. It worked well even with a small number of data in comparison to the number of models, i.e., 200 time-series and five models. 
To further validate this approach in an industry-setting, two comparisons were conducted between TDA and Dynamic Time Warping (DTW) with K-Means, see \cite{berndt1994using}. The first experiment consisted of 80 000 time series generated from the data set with added random noise. For TDA, it took less than 30 minutes on a standard commercial laptop, whereas DTW was not able to complete the process. A second experiment using DTW with K-Means was made under the same conditions described in \autoref{sec:tda}. It revealed that the first cluster consisting of AdaBoost, BSTS and BSTS classifier is the best suited for 69\% of the time series. For the second cluster containing Poisson regression and Random Forest, it was 31\%. Yet, it is still necessary to conduct experiments to 
verify the purity of the cluster.

\section{Conclusion}
This work had as an objective to provide practitioners with a system for demand forecasting consisting of preprocessing, training, and prediction of a large number of models as well as model selection. As a preprocessing technique, Diagonal Feeding was introduced. It helps demand planners improve the accuracy of their methods whenever future delivery dates are known and without requiring domain knowledge or extensive feature engineering. For prediction and testing, a large study comparing over fourteen methods was presented. Also, it applied a method based on matrix factorization for demand forecasting.
Similarly, a model selection method based on TDA was presented. In an industry-setting, low error metrics such as SMAPE can be misleading. The trained model might be incapable of forecasting the actual demand. The methodology provided alleviates this and shows better results than similar techniques while being easy to communicate to stakeholders. As a further line of work, this study would like to point out two main directions. First, for matrix factorization, there is the need to improve it for sparse data as well as to be more computationally efficient. Second, for the model selection based on TDA, it is worth considering different approaches not based on graph partitioning. One example is clustering based on point clouds. In conclusion, there is a need to up-skill existing personnel, and researchers can contribute to close this gap. Given the significant demand for analytics talent in the years to come, one can expect that the academic community will focus their attention in this direction.

\bibliographystyle{IEEEtran}
\bibliography{bib}

\end{document}